# High Brightness Lasing at Sub-micron Enabled by Droop-Free Fin Light-Emitting Diodes (LEDs)

Babak Nikoobakht,[1*] Robin P. Hansen,[1] Yuqin Zong,[1] Amit Agrawal,[1,2] Michael Shur,[3] Jerry Tersoff[4]

**Affiliations:**

[1] National Institute of Standards and Technology, Gaithersburg, MD 20899 USA.

[2] Maryland NanoCenter, University of Maryland, College Park, MD 20742 USA.

[3] Rensselaer Polytechnic Institute, 8th Street, Troy, NY 12180 USA.

[4] IBM T. J. Watson Research Center, Yorktown Heights, NY 10598 USA.

* Correspondence to: babakn@nist.gov

**Abstract:**

"Efficiency droop", i.e., a decline in brightness of LEDs at high electrical currents, has limited the performance of all commercially available LEDs. Until now, it has limited the output power of sub-micron LEDs and lasers to nanowatt range. Here we present a fin *p-n* junction LED pixel that eliminates efficiency droop, allowing LEDs brightness to increase linearly with injected current. With record current densities of 1000 KA/cm$^2$ (100 mA), the LEDs transition to lasing within the fin, with high brightness (over 20 µWatts). Despite a light extraction efficiency of only 15%, these devices exceed the output power of any previous electrically-driven sub-micron LED or laser pixel by 100 to 1000 times, while showing comparable external quantum efficiencies. Modeling suggests that spreading of the electron-hole recombination region in fin LEDs at high injection levels suppresses the non-radiative Auger recombination processes. Further refinement of this design is expected to enable development of a new generation of high brightness electrically addressable LED and laser pixels for macro- and micro-scale applications.

**One Sentence Summary:** A markedly bright fin light-emitting diode pixel is reported that with increasing electrical current flux can switch to a bright laser.

**Introduction:** Since the breakthroughs in GaN blue LED technology [1-3] and the subsequent realization of white LEDs, great progress has been made in wide bandgap (WBG) semiconductor LED lighting for general illumination,[4] display,[5] and many other applications[6] including biological sensing, surface disinfection and sterilization,[7] visible-blind photodetectors,[8] acoustic-optoelectronics,[9] and terahertz electronics.[10] In emerging miniaturized applications, these materials are heavily investigated for nanoscale LEDs and lasers for on-chip optical communication, [11] chemical sensing, [12, 13] high definition displays,[5] and visible light communication[14] to name a few. Impressive developments have been achieved in electrically-pumped nanoLEDs and low threshold nanolasers based on photonic crystals,[15, 16] nanopillars,[17, 18] or nanowires [19, 20] with output powers in the pico- to nanowatt range.[11] However, such nanoscale light sources still have too low power for most practical applications.[11] Like all LED technology, they are limited by "efficiency droop", which is the decline in internal quantum efficiency (IQE) with increasing current density.[21] In addition, there is a thermal decrease in IQE with increasing current due to heat generation at the junction. In planar LEDs, the IQE is shown to decrease by 30% as junction temperature increases from 23 °C to 177 °C.[22-24] As a result of these effects, measurements in nanoscale LEDs show a rollover of the IQE at injected currents as low as 0.3 mA to 4 mA. [17, 25-27]

The mechanisms causing droop are among the most heavily investigated topics in WBG materials. Efficiency droop has been linked to Auger recombination,[28, 29] phase space filling,[30] stimulated emission,[31, 32] delocalization-activated nonradiative recombination,[33-35] and incomplete carrier localization.[36, 37] Here we report a *p-n* heterojunction ZnO-GaN fin LED that inherently does not show efficiency droop even at record high current densities. Direct measurements and modeling show that the carrier loss mechanisms due to non-radiative pathways stay well controlled, which we attribute to the fin architecture. Measurement of the total spectral radiant flux shows that output power of the fin LED pixels increases linearly with drive current. Results show that the charge carrier radiative recombination efficiency is high enough to cope with the optical loss and allow lasing in the fin cavity at notably high current densities and temperatures.

**Results and discussion**

The fin LED architecture shown in figure 1A includes a lateral ZnO nanofin (1) epitaxially grown on *p*-GaN. ZnO has a wurtzite crystal structure and a bandgap ($E_g$) of 3.36 eV, similar to that of GaN ($E_g = 3.43$ eV). However, it has a considerably larger exciton binding energy of about 60 meV[38] (relative to 24 meV in GaN) making it more effective for high temperature light sources and optoelectronics resistant to radiation damage. The fins are grown via an Au-catalyzed vapor-liquid-solid (VLS)-growth process that is directed on the surface,[39] where the Au nanodroplets formed at the periphery of the catalyst pattern nucleate the nanofins (Fig. 1B). On *c*-GaN, a ZnO fin grows upright in the [0001] direction and laterally in the *m*-direction with an average width of about 160 nm and length of about 5 µm or longer. A ZnO fin could have one or two large non-polar ($11\bar{2}0$) side facets.[39] Fabrication of a fin LED is completed by connecting the *n*-type fin to the *n*-metal electrode. We used a "facet-selective contact" method (Fig. 1C-D), in which one of the fin side facets is first passivated with a dielectric layer (2), deposited at an angle (Fig. 1D). The other non-polar side facet remains accessible to receive the *n*-contact metal electrode (3), as presented in figure 1E. The side-facet selective contact method also results in the facile formation of clean metal-semiconductor interfaces as shown in the cross-section of figure 1F. The fabrication strategy that allows for large-area low-resistance contacts to fin LEDs, will be described elsewhere.[40]

Figure 2 presents the optical image (A) of a single fin LED pixel (under operation) and its magnified scanning electron microscope (SEM) image (B) delineating different device parts such as metal contacts. Electrically addressable fin LED pixels are fabricated using standard photolithography. Figure 2C shows an example of a linear array. In this optical image, fin LED pixels containing 1 to 10 fins are spaced 120 µm apart across a 1 cm range. For better current spreading in the *p*-GaN layer, the *p*-contact microelectrodes (yellow highlights) are spaced about every 170 μm and 300 μm.

Fin LED pixels are so bright that the individual pixels in the wire-bonded linear array can be directly measured in front of a metrology-grade charge-coupled device (CCD)-array spectroradiometer with no focusing optics (Fig. 3A). Using this setup, the total spectral radiant fluxes of individual fin LED pixels were measured at different drive currents. Figure 3B shows a series of representative electroluminescence (EL) spectra obtained from a pixel containing eight nanofins (pixel#17), with an average fin length of 5 μm and height of 1 μm. Two peaks are visible, at 3.37 eV (368 nm) and 3.28 eV (378 nm), as current is increased. These wavelengths agree with the neutral donor bound exciton ($D^0X$) recombination in ZnO. Observation of these two peaks at room temperature and their agreement with the room temperature cathodoluminescence (CL) data (filled spectrum) highlights the high quality of the ZnO fins and underlines the low concentration of the donor states relative to that typically observed in unintentionally *n*-doped ZnO nanowires.[38, 41] A low concentration of the donor states is also evidenced by the flow of holes from *p*-GaN to the *n*-ZnO and their recombination in the ZnO fin. Fitting these spectra gives a full-width-at-half-maximum (FWHM) of only 5 nm and 18 nm for these two peaks, demonstrating the low defect density of the fins. In the EL spectra, at lower injection currents, the appearance of the red shoulder at about 2.96 eV (420 nm) indicates radiative *e-h* recombination in GaN, due to the slow movement of holes toward the *p-n* interface. As the drive current increases, the intensity of the ultraviolet (UV) EL peaks at 3.37 eV (368 nm) and 3.28 eV (378 nm) continues to rise while the *e-h* recombination in GaN is suppressed.

To further explore the impact of drive current on the output power and spectral properties of fin LEDs, the total spectral radiance flux (W/nm) of a pixel (#50) containing five fins was investigated up to 100 mA (the limit of the measurement setup) using the integrating sphere described in figure 3A. The pixel was operated under a direct current (DC) bias during the operation times of 50 s, 25 s, 10 s, 5 s, 2.5 s, 1 s and 0.5 s and for the current range of 1 mA, 2 mA, 5 mA, 10 mA, 20 mA, 50 mA, and 100 mA, respectively. After each measurement, a 10-minute rest time ensured that the fin LED reached the equilibrium temperature of 25 (± 0.1) °C established by a temperature-controlled stage. The total radiant flux-current graph of figure 3C (dark circles) shows a nearly linear increase in the output power. For instance, 100 times increase in the injected current results in an about 108 times increase in the optical power for the pixel, i.e., 1 μWatt at 100 mA. This observation indicates that at high currents, the carrier loss due to non-radiative events is well controlled. The loss of carriers inside the active region at high injected currents is one of the main processes that promotes the efficiency droop. As the current increases, the rise of the intense UV EL peaks in the range of 3.37 eV to 3.18 eV (368 nm to 390 nm) in figure 4A shows that the radiative recombination occurs in the ZnO fin, as shown in the band structure of the ZnO-GaN heterojunction (Fig. 4B). In contrast to other types of LEDs, at high injection currents, the flow of electrons (electron leakage) to the *p*-GaN is not observed in fin LEDs. Such a leakage is an important factor in promoting the droop effect at high current densities.[42]

As current is raised, the EL UV peaks redshift in an approximately linear fashion (see Fig. 4c). For instance, the 3.37 eV (368.5) nm peak redshifts about 27 meV (3 nm) at 10 mA, and 144 meV (17 nm) at 100 mA, reaching 3.22 eV (385.4 nm). This shift is due to the junction temperature increase. The temperature dependent EL studies of the fin LEDs also show a linear redshift in the UV peak position (supplementary materials, Fig. S1). Based on these measurements, the 144 meV redshift in the fin LED corresponds to a drastic rise in the junction temperature to approximately 340 °C during its 0.5 s operation. At the low injection current of 2 mA, the temperature dependent EL analysis from 25 °C to 110 °C (Fig. 4D) shows no significant variation in the charge recombination in the ZnO fin. Hence, the defect-related Shockley–Read–Hall (SRH) non-radiative carrier recombination in the fin is minor. Meanwhile, as the as temperature is raised, a gradual suppression of *e-h* recombination in GaN is observed (supplementary materials, Fig. S2). These results underline the tolerance of ZnO fin LEDs to non-radiative recombination processes both at low and high current densities. The resilience of fin LEDs to high temperature is also in line with a higher exciton binding energy of ZnO compared to GaN.

The fin LED pixels with the fin numbers smaller than five, consistent with pixels with more fins, show that their total radiant flux (output optical power) grows linearly with injected current and at 50 mA reaches 1 µW to 3.5 µW, as shown for 4 randomly selected pixels in figure 4E. Taking into account the active area of a fin LED, these values correspond to output power densities ranging from 16 W/cm$^2$ to 235 W/cm$^2$ (supplementary materials). It is noteworthy that at 50 mA, fin LED pixels receive substantial current densities in the range of 0.45 MA/cm$^2$ to 2.2 MA/cm$^2$ depending on the number of their fins (supplementary materials), which highlights their advantageous large side facets for current injection. Nanowire-based LEDs are shown to receive current densities in the range of 0.2 kA/cm$^2$ to 7 kA/cm$^2$, which, typically, places them in their droop regime.[26, 27]

Full three-dimensional Finite-Difference Time-Domain (FDTD) modeling (Fig. 4F) shows that the fin LEDs emit light from their facets open to air and have an intrinsic light extraction efficiency (LEE) of about 15 %. The rest of the light is trapped in the *p*-GaN substrate, due to a higher refractive index of GaN (2.5) relative to ZnO (2.2). The measured range of output power densities in figure 4E indicates that fin LEDs have the potential to exceed the power density of commercial planar UV-A LEDs of ≈ 23 W/cm$^2$ to 75 W/cm$^2$ that are engineered to have greater than 85% light extraction efficiencies.[21, 43] By increasing the LEE beyond 15%, we estimate up to five times increase in the brightness of fin-based light sources, for instance, by coupling them to waveguide modes or reducing the light trapping in the GaN slab. Reports on power measurements of individual nanowire LED pixels is scarce; however, recent measurements using ensembles of free-standing nanowire LEDs show power density values in the range of 0.001 W/cm$^2$, 0.5 W/cm$^2$, and 3 W/cm$^2$, respectively, for axial InGaN/GaN nanowires,[44] InGaN/AlGaN dot-in-a-wire core-shell nanowires,[45] and core-shell AlInGaN nanowires.[46]

As the current density reaches above 50 mA (at about 500 KA/cm$^2$) in some of the fin LED pixels, the broad EL emission at 385 nm narrows to two intense lines at 403 nm and 417 nm. This is a clear indication of lasing. A representative example, presented in figure 5A, shows the spectral evolution of pixel #43 at different drive currents and appearance of the two sharp peaks at 100 mA. Similar lasing is seen in other pixels (Fig. S3). In pixel #43 we observe a linear redshift in the EL peak positions as current is raised, similar to pixel #50 (Fig. 4C), which we attribute to the rise of the junction temperature (Fig. 5B-i). We note that pixel #43 with one fin shows 143 meV (16.5 nm) redshift in the EL peak at 50 mA, suggesting the junction temperature of approximately 340 °C. The total radiant flux of the pixel #43 during LED and lasing emissions, presented in figure 5B-ii (red squares), shows the record output powers of 1.9 µW and 20 µWatts at 50 mA and 100 mA, respectively. The FWHM of the EL emission in figure 5B-ii (purple diamonds), narrows from 65 nm to 2.7 nm once the device is pumped at 100 mA. This is 10 times increase in the output power due to a two-fold increase in the injected current. Figure 5C-i compares the output power of this pixel (red circles) operating as a laser diode with other pixels as LEDs. If pixel #43 were an LED, we would have expected only about a two-fold increase in the output power as seen in pixel#50 of figure 5C-i (purple triangles). Interestingly, similar to pixel #43, we observe an excellent consistency in lasing emission lines in other lasing pixels (such as pixels #28 and #49, Fig.S3).

Our results suggest that a ZnO fin on GaN can act as a Fabry–Pérot cavity. Numerical modeling using three-dimensional FDTD technique show the corresponding spectral response as measured in the far-field when excited by a randomly oriented broadband dipole located at the interface between ZnO and GaN (Fig. S4). Various passive cavity modes separated by the free-spectral range (FSR) of approx. 14 nm to 16 nm the cavity is consistent with the experimentally measured spacing between the two lasing modes (14 nm). Note that the experimentally observed FSR of 14 nm corresponds to an approx. cavity length of 2.5 µm (assuming a mode index of 2.4). The typical FWHM of the passive cavity modes is approx. 9 nm corresponding to a quality factor ($Q = \lambda / \delta\lambda$) of the resonant mode of ≈ 45. This factor is limited primarily by the radiative losses (to free-space and waveguiding in the GaN layer). By patterning the GaN layer around the nanofin, we except the quality factor to improve by an order of magnitude. Upon electrically pumping the cavity above the lasing threshold, the two lasing modes at 403 nm and 417 nm appear (Fig. 5A). The decrease in linewidth from 9 nm to 2.7 nm upon lasing is consistent with that expected of a transition from the spontaneous emission to the stimulated emission regime. We believe that the lasing linewidth in the current configuration is both limited by the radiative losses (as discussed above) and by the non-radiative recombination due to high junction temperature. The ability of a fin with an approximate volume of 0.8 μm³ to reach population inversion is quite striking as droop effects and optical loss were expected to overwhelm the fin cavity for the junction temperature of about 340 °C and 1000 KA/cm² current densities. Previous theoretical modeling has also suggested that the thermal effects can make nanoLEDs and nanolasers operation unsustainable beyond 100 KA/cm².[11]

To better understand the impact of e-h recombination pathways in fins, we investigated the wall-plug efficiency (WPE) of the fins in LED and lasing modes. Figure 5C-ii shows the general trend in WPE for pixels in LED mode that has a rise followed by a decline, while for the lasing pixel (red circle) the WPE suddenly increases at 100 mA. The WPE ($\eta_{WP}$) of a fin LED is $\eta_{WP} = \eta_{INJ} \times \eta_{LEE} \times \eta_{IQE} \times \eta_{DRIVER}$, respectively, including current injection, LEE, IQE and driver (feeding) efficiencies. The driver efficiency, $\eta_{DRIVER} \sim \frac{E_{gap}}{qV_{bias}}$, is the ratio of the mean energy of the photons emitted and the total energy that an electron-hole pair acquires from the power source. This value was calculated using the voltage-current scans of the fin LEDs (Fig.S5) and used in the WPE equation above to extract the IQE dependences of the fin LED-laser pixels as illustrated in the semi-log graph of figure 6A. As seen, the IQE of fin LED pixels do not show the efficiency droop even at very high current densities of 1000 KA/cm². Furthermore, the IQE shows an abrupt increase when the pixel becomes a laser diode. This is remarkable, as normally LEDs show a strong decline in their IQE efficiency as current density increases. Experimentally measured WPE or EQE values for nanowire LEDs based on absolute radiant flux data are scarce. Earlier semi-quantitative estimations in GaN based nanowire LEDs show that EQE or IQE peaks at small injection currents ranging from 0.4 mA to 5 mA[25-27, 47] corresponding to current densities of 0.2 KA/cm$^2$ to 10 KA/cm$^2$. A more recent quantitative report on metal-cavity InP-based nanopillar LEDs, shows that about 22 nanowatt power at max EQE of 10$^{-4}$ is reached at a current range of 0.2 mA to 0.4 mA (current density of 100 kA/cm$^2$).[17] Since the fin design prevents the roll over in EQE, it is possible to achieve a comparable EQE of 6 × 10$^{-5}$, but at 40 times higher current densities. This advantage results in fin pixels with 1000 times more output power than that of the nanopillar LED design with one of the best reported performance.

As we discuss below, the heterodimensional fin architecture is conducive to limiting the droop effect and, therefore, helps boosting IQE, which is a determining parameter in the overall performance of the LED. Another key factor for achieving high power emission and lasing is the high quality of ZnO crystal evidenced by the narrow bandwidth transitions observed by the CL and EL spectroscopies. To support this interpretation, we present a three-parameter model based on the radiative and nonradiative processes (ABC) including the SRH and Auger recombination,[48] in which the energy is eventually released as heat. According to the ABC model (supplementary materials, Eq.1), the IQE is given by:

$$\eta = \frac{N_p}{1 + N_p + c_a N_p^2}. \qquad (1)$$

Here $N_p = n/n_o$, is the dimensions electron-hole pair density, $n_o = \frac{A}{B}$, $c_a = CA/B^2$, and $c_a$ is the dimensionless Auger recombination constant. This model (Fig. S6) describes the impact of rise of non-radiative Auger recombination and decline of IQE. In the steady state, the current density flux gradient is estimated as $J = I/(qd)$, where $J = An + Bn^2 + Cn^3$. Here $I$ is the current density, $q$ is the electron charge, and $d$ is the characteristic length of the recombination region. The current density flux gradient ($J$) could be rewritten as $j = N_p + N_p^2 + c_a N_p^3$. Here, the dimensionless current flux density is $j = J/(An_o)$.

We consider two limiting cases: (1) the low injection regime when the generation occurs close to the heterointerface. In this case, the Auger recombination is negligible. (2) High injection regime, when the generation occurs in the fin and the Auger recombination is dominant. In the low injection case, we neglect the Auger recombination contribution in the current flux density. In this case, the SRH defect-related carrier recombination is more dominant. However, it does not cause the efficiency droop. In the high injection regime, the Auger recombination is dominant, and the equation for the current flux density becomes $j = N_p + N_p^2 + c_a N_p^3 = c_a N_p^3$. In this regime,

$$\eta_{high} = \frac{1}{c_a^{2/3} j^{1/3}} \tag{2}$$

This equation applies when $c_a N_p \gg 1$ or $(jc_a^2)^{1/3} \gg 1$. The general expression for IQE could be interpolated as

$$\eta = \frac{1}{1/\eta_{high} + 1/\eta_{low}} \tag{3}$$

The key difference between the conventional LED and the fin LED is the dependence of current gradient $j$ on the injection level. For conventional LEDs, the length of the recombination region, $d$, in the steady state current density flux gradient $J = I/(qd)$, is nearly independent of injected current, $I$. For the fin LEDs (Fig. 6B), $d$ increases with $I$ and the current gradient $j$ saturates leading to the saturation of the Auger recombination. For the conventional LED, we obtain the dependences shown in figure 6B. In this graph, as the Auger recombination constant $c_a$ increases, the IQE declines (top to bottom curves). For the case of the fin LED, we assume that at high injections the electron hole pairs spread out, *i.e.*, $d$ increases (Fig. 6B). The reason for the recombination region expansion into the fin is the heterodimensional bottleneck effect occurring at the $n$-ZnO fin/$p$-GaN heterointerface. Since the current in the narrow fin must be equal to the current in the *p*-GaN region, the recombination length in the fin has to expand with the rising current in order to provide enough electrons to recombine with the holes collected from a much larger area. Consequently, the current gradient remains nearly constant with the rising current. The mathematics describing this current distribution is analogous to the problem of calculating the depletion region length for a heterodimensional (three dimensional to two-dimensional) p-n junction.[49] According to this model, the lasing occurs when the recombination region reaches the top of the fin and no longer expands with an increase in the injected current. For the high injection regime of a fin LED, Eqs. (2) and (3) predict nearly droop-free dependences shown in Fig. 6D.

We argue that both the SRH and Auger recombination pathways are mitigated in the ZnO fin architecture. At the low injection regime, when the generation occurs close to the heterointerface, the non-radiative SRH in ZnO is insignificant (Fig. 4D). Output power in Figure 5C also show that at high injection regime, when the generation occurs in the ZnO fin, the output power grows linearly with the injected current indicating carrier loss due Auger recombination is minimal. The presented mechanism highlights the significance of the two factors stated above in improving the IQE at high current densities. Namely, the fin shape facilitating the current gradient saturation and the high quality of ZnO crystal. In addition, ZnO fins do not host the threading dislocations originated at the underlying GaN substrate. This is due to their growth mechanism that governs the lateral growth in the surface-directed VLS process versus the epitaxy used for conventional planar LEDs. The ZnO-GaN thin film-based LEDs have shown the efficiency droop effect.[50] Hence ZnO is most likely not the critical compound for creating a droop-free LED. A ZnO fin, due to its sub-200 nm width is expected to have a lower potential drop and resistive loss across its width (the two factors affecting the IQE). The comb like structure of the *p*-GaN contact is also effective for the current spreading and for the effective hole injection into the fins.

Overall, the linear rise of the output power versus the injected current at high current densities shows the effectiveness of the fins in charge injection. Furthermore, the minimal impact of temperature on the defect-related SRH carrier loss in ZnO fin, at low injection current underlines the low defect density of the fins. There is no electron-blocking layer in the presented fin heterojunction. Nevertheless, there is no detectable electron leakage and/or electron hole recombination in the *p*-GaN at high current densities. Our results strongly indicate the absence or negligible presence of the critical factors that reduce the efficiency of conventional LEDs as well as sub-micron LEDs including electron leakage, Auger recombination, defect-related recombination and temperature effects.

**Conclusions**

LEDs have been considered as the ideal lighting sources for a vast range of applications including residential, industrial and automotive lighting, smart lighting, displays, Visible Light Communications, biomedical applications, water and food sterilization, security systems, and sensors, to name a few. Miniaturized LED and laser pixels with high-luminance/high-radiance point sources will revolutionize many of these applications including the advanced smart light sources in far-field applications such as multi-pixel LEDs and intelligent selective pixel control for use in smart lighting, display technology, and internet of things. NanoLEDs and nanolasers with a strongly reduced cavity volume have the potential to reach much higher modulation speeds for a wide range of near-field applications in sensing, metrology, microscopy, and on-chip communications. The main remaining challenge is that the existing nanoLEDs and nanolaser are too dim for most practical applications, with output powers in the nanowatts range, due to the dominance of temperature- and current density-droop effects. The architecture reported here is a simple fin ZnO-GaN LED with a sub-micron area that alleviates or removes these limitations. This design allows generation of output powers in the microwatt range, about 1000 times higher than that of existing nanoscale LEDs and lasers. We relate this enhanced performance to the impact of fin shape on mitigating the non-radiative pathways and providing large side facets for an effective electrical current injection and forming a laser cavity. Evidence of lasing in fin cavities with an estimated temperature well above room temperature further underlines the droop-free nature of the fins at extreme current densities. The findings here are highly encouraging and represent a significant step in designing architectures that can overcome the "current density droop" and "temperature droop" effects in LEDs, to maximize their achievable output power per pixel. While the reported nanoLEDs and nanolasers operate in the near UV range, this concept could be applied to the materials system, such as AlGaN, BN or to their heterostructures to develop far brighter deep UV LEDs and lasers. At microscale, these findings are expected to enable further reduction in the cavity size below sub-100 nm while offering high-performance and high-brightness light sources.

**Materials and Methods**

**Preparation of n-ZnO/p-GaN heterojunctions**: The substrate for the lateral fin growth included a 200 nm- thick Mg-doped *p*-GaN epi-layer on *c*-plane sapphire with an acceptor concentration of 5 x $10^{17}$ cm$^{-3}$ (Kyma Technologies[*]). The *p*-GaN epi-layer was grown on a 1500 nm, undoped GaN layer. GaN wafers were diced with a dicing saw to the one square cm dies. The wafer cleaning steps included solvent wash including hot acetone followed by isopropyl alcohol and drying with nitrogen gas.

*Gold catalyst micropatterning*: One cm$^2$ *p*-GaN substrates were photopatterned with a Carl Suss Mask aligner[*] with a nominal resolution of 2 μm. The Shipley 1813 photoresist layer was spin casted according to established clean room protocol. This layer was spun on the surface at 66.7 Hz (4000 rpm) in 45 s. After a one-minute bake at 115 °C, it was patterned with catalyst micropatterns.

---

[*] Certain commercial equipment, instruments, materials, or software are identified in this paper to foster understanding. Such identification does not imply recommendation or endorsement by the National Institute of Standards and Technology, nor does it imply that the materials or equipment identified are necessarily the best available for the purpose.

The 10 nm to 20 nm thick gold patterns were deposited using a thermal evaporator. Photoresist lift off was carried out in hot acetone followed by wash in warm and cold Isopropyl alcohol.

*Growth of ZnO nanofins*: ZnO nanofins were grown on gold-patterned *p*-GaN surface using a modified surface-directed vapor-liquid-solid growth process reported elsewhere.[40] Growth was carried out in a horizontal tube furnace with 800 mm length and 49 mm inner diameter. The ZnO/Graphite mixture of 0.140 g (1:1 mass ratio) was positioned at the center of a small quartz tube with a 130 mm length and 19 mm inner diameter. The tube furnace temperature reached 890 °C (with a ramp rate of ≈111 °C per minute) and a dwell time of 40 minutes under ≈0.5 standard liters per minute (SLPM) flow of ultra-dry (99.99 %) $N_2$ gas.

**Measurement of optical power of fin LED and laser pixels:** The fin LED chips were mounted on a temperature-controlled mount (TCM) through a metal-core printed circuit board (MCPCB) (Fig. 3a). The fin LEDs were measured for total spectral radiant flux (W/nm) using a metrology-grade CCD-array spectroradiometer with a spectral range from 300 nm to 1100 nm and a bandpass of 2.5 nm. The input optic of the spectroradiometer is the irradiance probe composed of a 44 mm diameter integrating sphere with a 16 mm diameter entrance port and a 5000 mm long quartz optical fiber bundle with 1.5 mm diameter core. The spectroradiometer was corrected for spectral stray light errors using the method described elsewhere.[51] Prior to the measurement of the fin LEDs, the spectroradiometer was calibrated for spectral irradiance responsivities. The calibration used a 1000 W spectral irradiance standard FEL lamp at 500 mm, and the spectroradiometer responsivity was converted to total spectral radiant flux responsivity by multiplying the spectral irradiance of the FEL lamp.

The fin LED chip was placed at the center of the entrance port and was less than 5 mm from the 16 mm diameter entrance port so that the integrating sphere could collect more than 95 % of light emitted by the LED. The errors resulting from the interreflection between the MCPCB of fin LEDs and the 16 mm diameter entrance port of the integrating sphere was estimated to be less than 5 %. The fin LEDs were measured for total spectral radiant flux at different operating current levels and the spectroradiometer's integration time was adjusted accordingly; ranging from 65 s to 0.02 s. The measured total spectral radiant flux (W/nm) was used to calculate the total radiant flux (W) covering the spectral range from 350 nm to 800 nm. The temperature of the TCM, where the device was mounted, was set to 25 °C for all the measurements. The ten minutes intervals between the consecutive measurements allowed the fin LED to reach its equilibrium temperature.

**Acknowledgments**

Authors would like to thank Christopher B. Montgomery at the Biomolecular Measurement Division of NIST for his kind support and performing the wire bonding related to this project. A. A. acknowledges support under the Cooperative Research Agreement between the University of Maryland and the National Institute of Standards and Technology, Center for Nanoscale Science and Technology, Award#70NANB14H209, through the University of Maryland. The work at RPI was supported by the US Army Cooperative Research Agreement.

**Author contributions**

B.N. conceived and oversaw the experiments. Y.Z. designed the EL measurement setup. Y.Z. and B.N. conducted the EL measurements. R.P.H. fabricated the LED device. M.S. provided the theoretical model. A.A. performed the optical modeling. J.T. assisted in analyzing data and improving the fin growth. B.N. wrote the manuscript with help of J.T., M.S. and other authors.

**Data and materials availability.** All data needed to evaluate the conclusions in the paper are present in the paper and/or the Supplementary Materials. Additional data available from authors upon request.

**Competing interests**

The authors declare no competing interests.

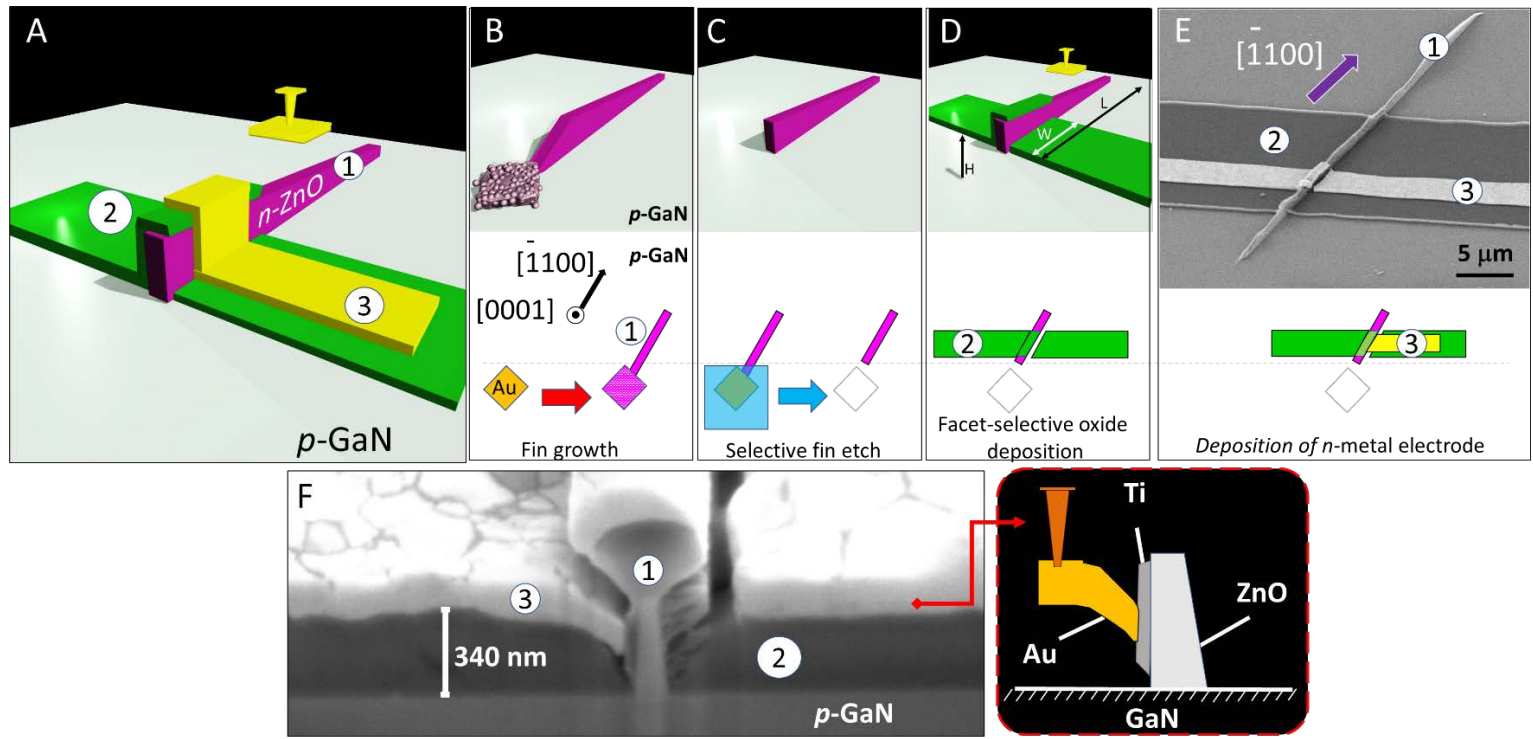

**Figure 1. Architecture and fabrication steps of *n*-ZnO Fin LED on *p*-GaN.** (A) A fin LED pixel includes: *n*-ZnO fin (1), isolating dielectric material (2) and *n*-metal contact (3). (B) Surface-directed ZnO fin epitaxially grown on *c*-plane GaN from an Au catalyst pattern via surface-directed VLS growth process. (C) Isolated fins are formed by removing the initial catalyst site via photolithography and wet etch. (D) One of the fin side facets is passivated via photolithography and angled-oxide deposition. The open facet has an approximate area of H x W, where "H" is height and "W" is width. The maximum length of W can be the fin length (L). (E) The open side of the fin receives the *n*-metal contact. (F) Electron micrograph of cross-section of a fin LED. Inset schematically shows the fin, *n*-metal contact, and GaN substrate.

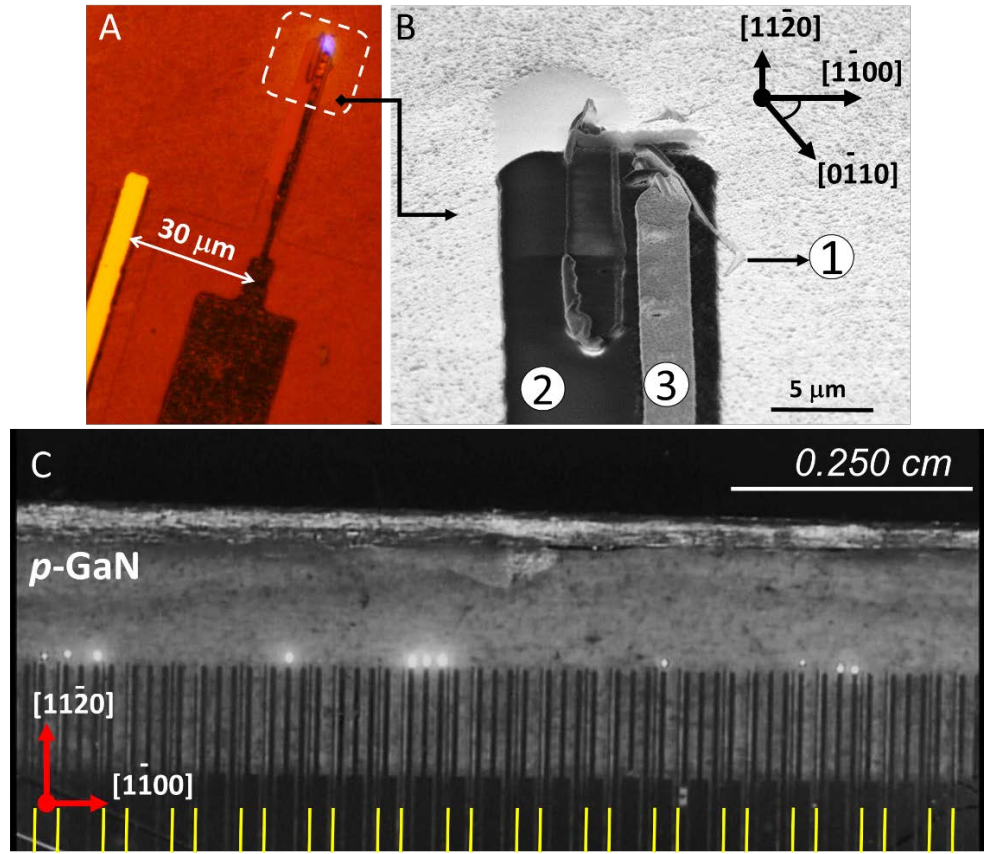

**Figure 2. A Fin LED pixel.** (A) Optical image of a LED pixel containing a single fin. The yellow electrode (true color) in the left side is the *p*-metal contact. The marked area shows the fin LED under forward bias. Its SEM image is shown in part (B) highlighting different layers of the pixel and the GaN crystallographic directions. (C) Optical image of a linear array of fin LED pixels. The bright spots represent fin LED pixels that are ON. The yellow lines highlight the position of *p*-contact electrodes relative to the *n*-contacts. In the image the ZnO fins grow direction is in $[1\bar{1}00]_{GaN}$.

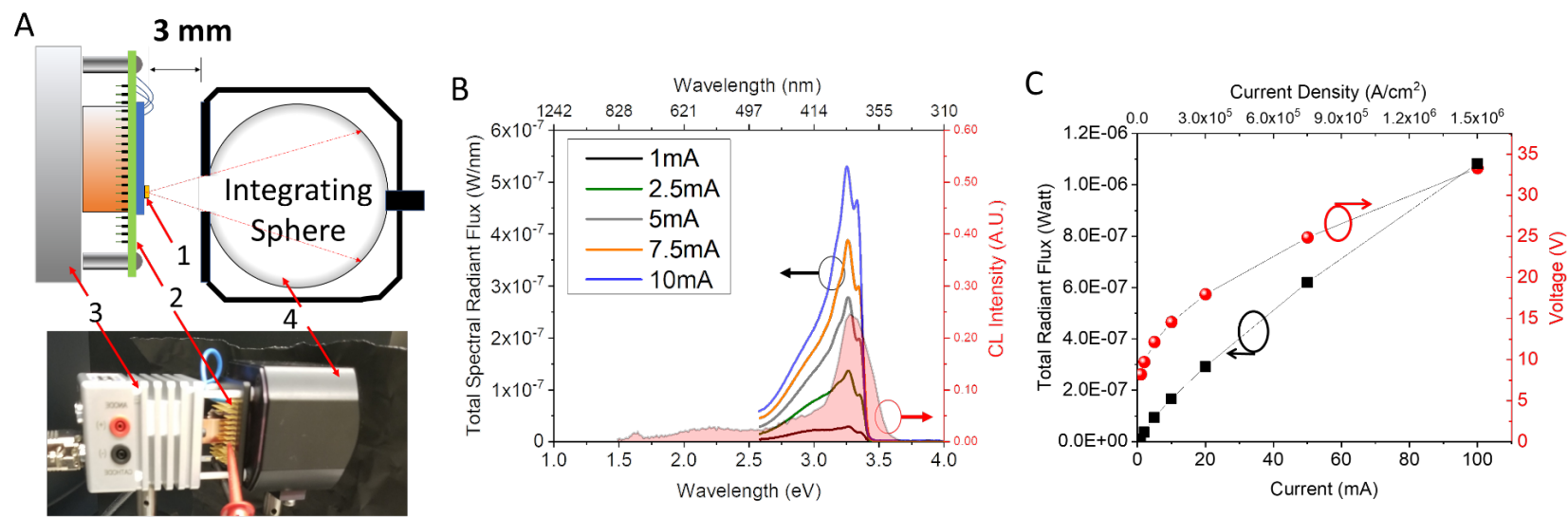

**Figure 3. Electroluminescence and output power of fin LEDs.** (A) Wire-bonded array of fin LEDs mounted in front of a 44 mm diameter-integrating sphere, which is connected to a calibrated metrology-grade CCD-array spectroradiometer using an optical fiber bundle. Fin LED pixel (1); chip carrier (2) mounted on a stage (3); detector (4). Photo Credit: Babak Nikoobakht, National Institute of Standards and Technology. (B) Total spectral radiance flux (W/nm) of a representative pixel containing eight fin LEDs as the injection current is increased from one mA to 10 mA. Each spectrum is an average of five measurements with a total time of 325 s. Two intense UV emission peaks appear at 368.5 nm (3.369 eV) and 378.5 nm (3.280 eV). The filled spectrum is the CL of a single ZnO fin at 2 keV. (C) Total output power (left axis) and voltage-current data (right axis) of a representative pixel containing three fin LEDs at different injection currents of 1 mA to 100 mA. Expanded uncertainty in power measurement is less than 5 %, with coverage factor of k=2.

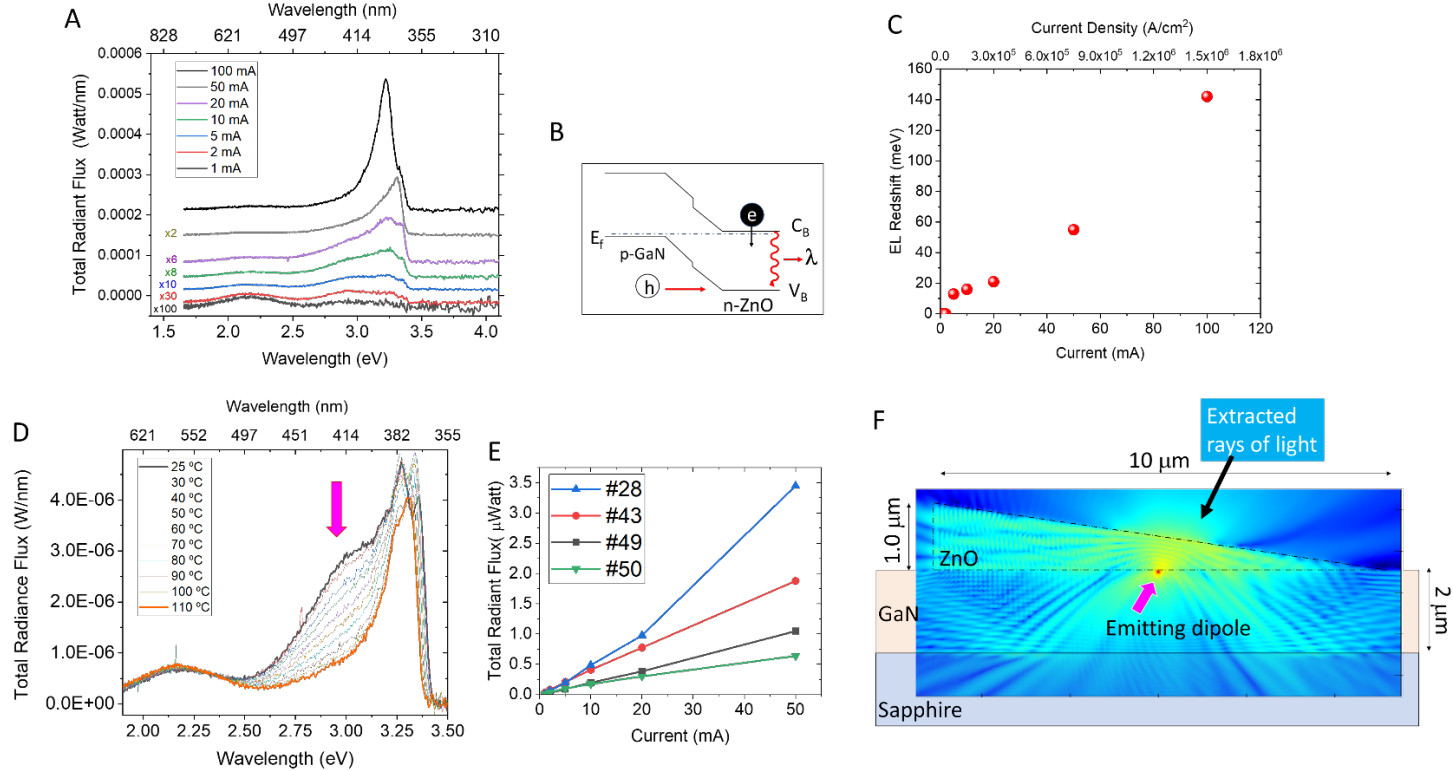

**Figure 4. Electroluminescence of fin LEDs at high current density and heat**. (A) Total spectral radiance flux (W/nm) of a representative pixel containing 3 fin LEDs as the injection current is raised from 1 mA to 100 mA. The acquisition time at each drive current is 50 s, 25 s, 10 s, 5 s, 2.5 s, 1 s and 0.5 s, respectively. (B) ZnO-GaN band structure. (C) UV peak position redshifts (in eV) as current is increased. The expanded uncertainty of the measured peak position is less than 0.5 nm, with coverage factor of k=2. (D) Change in EL of the fin LED pixel at a low current load of 2 mA as temperature increases from 25 °C to 110 °C using a temperature-controlled stage. The broad emission at 420.2 nm (2.955 eV) due to e-h recombination in the GaN side does not redshift, but is suppressed as temperature reaches 110 °C. The intensity of the first UV emission related to e-h recombination in the ZnO fin at 368.5 nm (3.369 eV) stays almost the same, but redshifts 4.8 nm and nearly overlaps with the second UV peak. (E) Total radiant flux (output power) of four randomly selected fin LED pixels in the range of 1 µWatt to 3.5 µWatt (± 10 %). (F) FDTD simulation of light extraction from a fin LED viewed from the long side of the fin. Dimensions are specified in the figure. A single dipole is oriented 45 degrees at the interface. Light rays leave more from the tall sides of the fin. Model shows about 8 % emission from sides and 7 % from top.

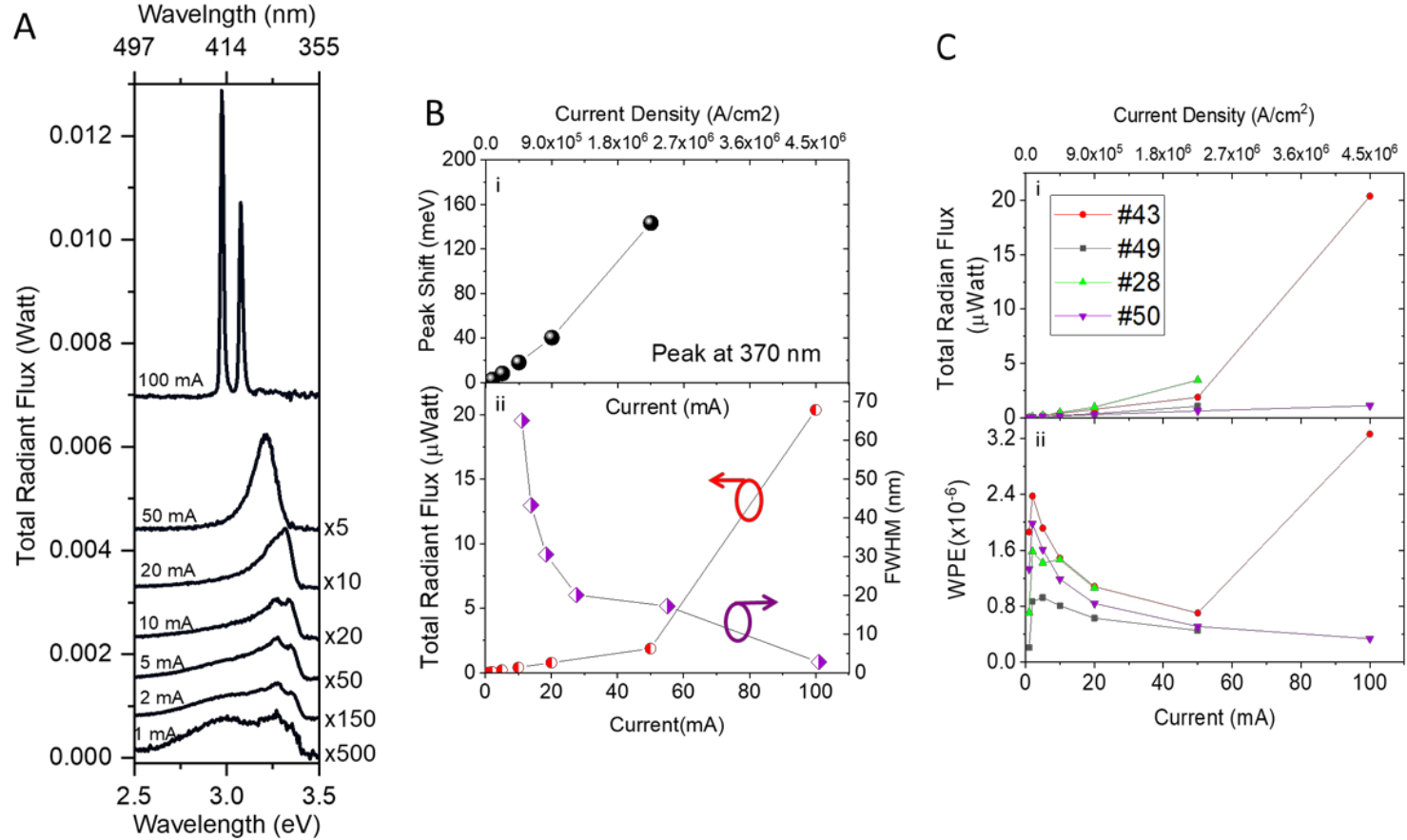

**Figure 5. Transition of fin LEDs to fin laser at high current density**. (A) Total spectral radiant flux of fin LED pixel #43 in the range of 1 mA to 100 mA shows the EL spectral evolution. The acquisition time at each drive current is 65 s, 30 s, 5 s, 1 s, 0.5 s, 0.1 s and 0.02 s, respectively. (B-i) While pixel #43 is in the LED mode, the shift in EL peak position varies linearly as drive current is raised to 50 mA. (B-ii) Total radiant flux, in microwatt, output power (left axis) and EL linewidth narrowing (right axis) of pixel #43 in the range of 1 mA to 100 mA. (C-i) Total radiant flux of pixel #43 (red), in microwatt, in LED and lasing modes and its comparison with three other LED pixels. (C-ii) WPE graphs of pixel #43 (red) in lasing mode and its comparison with three other LED pixels. In B and C, current density in the upper part of the graph is for pixel#43.

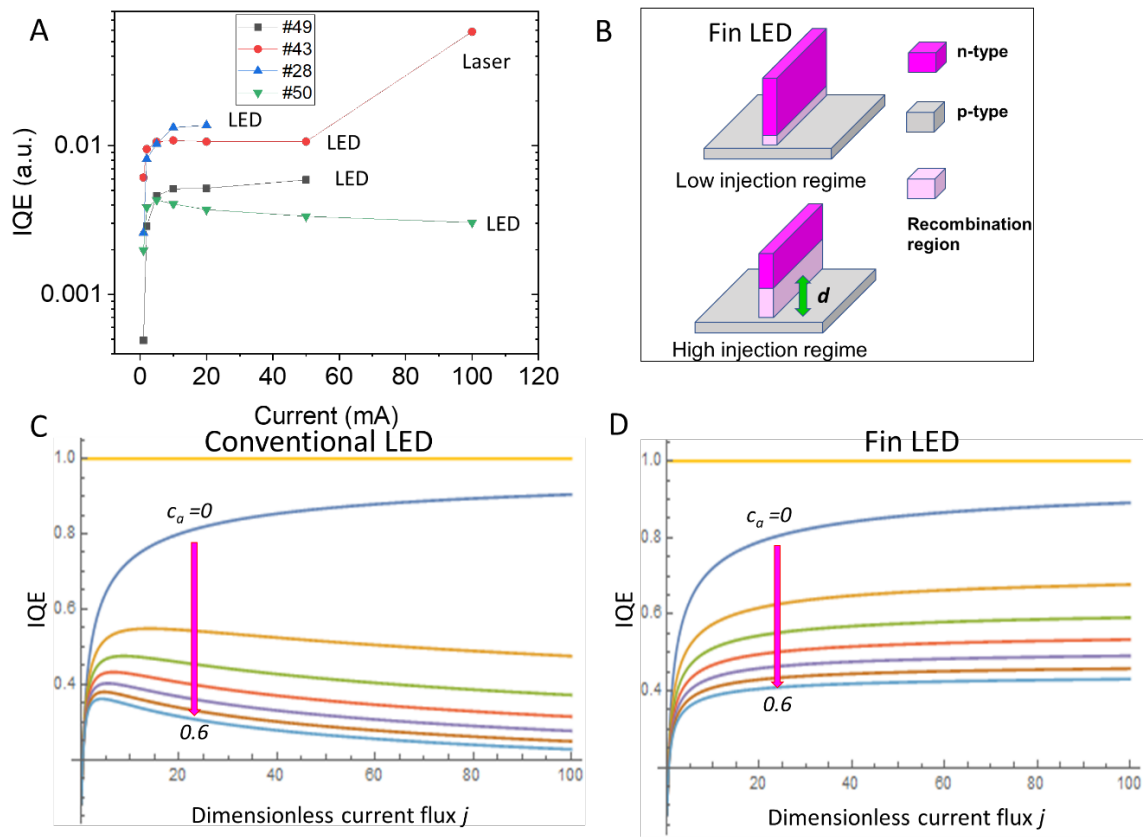

**Figure 6. IQE efficiencies of fin and planar LEDs.** (A) IQE graph of pixel #43 (red) in LED and lasing modes and its comparison with three other LED pixels. The top scale presents the current density for pixel#43. (B) Fin LED charge spreading length "d" at low and high current injection regimes. (C) Calculated IQE vs. dimensionless current flux $j$ for conventional LEDs, from model Eq. 3. The dimensionless Auger recombination constant $c_a$ varies from 0 to 0.6 (step 0.1). (D) Calculated IQE vs. dimensionless current flux $j$ for Fin LED, from model Eq. 5. The dimensionless Auger recombination constant $c_a$ varies from 0 to 0.6 (step 0.1). The parameter used in the calculation is the dimensionless injection current flux $j_o = 5$.

# High Brightness Lasing at Sub-micron Enabled by Droop-Free Fin Light-Emitting Diodes (LEDs)

Babak Nikoobakht,[1] Robin P. Hansen,[1] Yuqin Zong,[1] Amit Agrawal,[1,2] Michael Shur,[3] Jerry Tersoff[4]

**Affiliations:**

[1] National Institute of Standards and Technology, Gaithersburg, MD 20899 USA.

[2] Maryland NanoCenter, University of Maryland, College Park, MD 20742 USA.

[3] Rensselaer Polytechnic Institute, 8th Street, Troy, NY 12180 USA.

[4] IBM T. J. Watson Research Center, Yorktown Heights, NY 10598 USA.

* Correspondence to: babakn@nist.gov

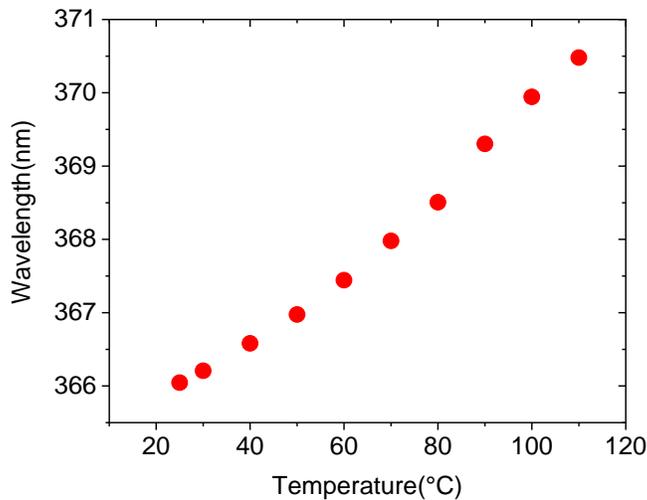

**Figure S1.** Redshift in UV EL peak (368 nm) of fin LEDs as the mount temperature was raised from 25 ˚C to 110 ˚C. The pixel contains 3 fin LEDs and is driven at 2 mA of DC current at each temperature. Blue shoulder of the peak was used for measuring the redshift. This method is commonly used for estimating the junction temperature in planar LEDs.

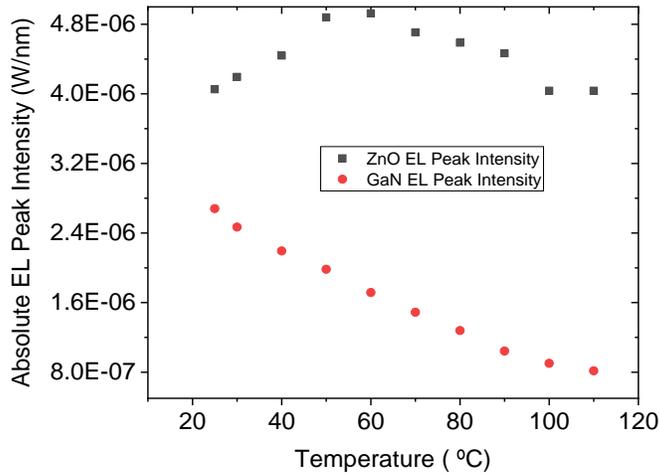

**Figure S2.** A representative pixel containing 3 fin LEDs driven at 2mA of DC current. Impact of temperature on intensity of EL peak for ZnO at 386 nm (dark points) and GaN peak at 420 nm. Temperature was changed from 25 °C to 110 °C. For ZnO, the emission intensity first shows an increase, and then reaches its initial value. For GaN peak, it declines as temperature is increased.

**1. Current density for pixels with different fins:**

For a fin with 1 μm height (and 5 μm length) that has received a *n*-metal contact with linewidth of 3 μm, the electrical contact deposited on the fin side facet (Fig. 1) for electrical injection is approximately 2.24 x $10^{-8}$ cm$^2$ (2.8 μm x 0.8 μm). Therefore, for dev. # 49 with three fins, in figure 5A, the 50 mA drive current corresponds to a current density of 740 kA/cm$^2$, while dev.#43 with one fin, receives 2.2 MA/cm$^2$.

**2. Calculation of output power density for pixels with different number of fins:**

The active area of a single fin is the base area of the fin, namely, its length (5μm) x width (0.16 μm). In table below, for every pixel, the active area was calculated based on its number of fins. To calculate the output power density, the total radiant flux (Watt) was divided by the total active area of a pixel. As examples, power densities for pixels containing 1 and 5 fins are shown in table below with two different colors.

| Pixel # | #fins | Fin area (μm2) 0.8 Total active (pn junction) area(cm2) | I (mA) | Output Power desnities 1 | 2 | 5 | 10 | 20 | 50 |
|---|---|---|---|---|---|---|---|---|---|
| #50 | 5 | 4.00E-08 W/cm2 | | 0.273131 | 0.962821 | 2.447438 | 4.32492 | 7.517024 | 15.93598 |
| #28 | 4 | 3.20E-08 W/cm2 | | 0.341068 | 2.005942 | 6.026625 | 15.18932 | 30.535 | 107.7971 |
| #43 | 1 | 8.00E-09 W/cm2 | | 3.288104 | 9.788895 | 25.83951 | 50.92718 | 96.82204 | 234.601 |
| #49 | 3 | 2.40E-08 W/cm2 | | 0.096019 | 1.029799 | 3.812658 | 8.156566 | 15.96752 | 43.61691 |

**3. Calculation of current density for pixels with different number of fins:**

The contact area of metal with fin is width of the metal microelectrode (2.79 μm) x height of the fin (0.8 μm). Depending on the number of fins per pixel, this area would change. For instance, at 50 mA, fin LED pixels with 1 and 5 fins experience current densities of 0.449 MA/cm$^2$ to 2.2 MA/cm$^2$ as shown in table below with two different colors.

| Pixel # | #fins | Totcal Contact-area(cm2) | I (mA) | Current densities | | | | | |
|---|---|---|---|---|---|---|---|---|---|
| | | | | 1 | 2 | 5 | 10 | 20 | 50 |
| #50 | 5 | 1.116E-07 | | 8960.573 | 17921.15 | 44802.87 | 89605.73 | 179211.5 | 448028.7 |
| #28 | 4 | 8.928E-08 | | 11200.72 | 22401.43 | 56003.58 | 112007.2 | 224014.3 | 560035.8 |
| #43 | 1 | 2.232E-08 | | 44802.87 | 89605.73 | 224014.3 | 448028.7 | 896057.3 | 2240143 |
| #49 | 3 | 6.696E-08 | | 14934.29 | 29868.58 | 74671.45 | 149342.9 | 298685.8 | 746714.5 |

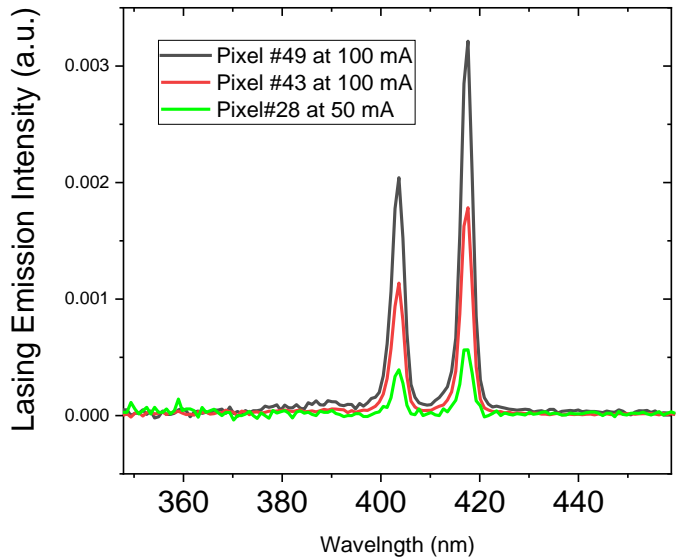

**Fig.S3.** Consistency of the lasing wavelengths in three different pixels.

**4. Calculation of spectral response of the fin cavity:** The spectral response of the cavity formed by a cuboid shaped ZnO nanofin (length 5 µm, width 160 nm and height 1 µm) placed on top of a 2 µm thick GaN layer on a sapphire substrate is numerically modelled using three-dimensional finite-difference-time-domain technique. The corresponding spectral response as measured in the far-field when excited by a randomly oriented broadband dipole located at the interface between ZnO and GaN is shown in Fig. S4. Various passive cavity modes, separated by the free-spectral range (FSR) of the cavity, are clearly evident. Spacing between the modes of approx. 14 nm to 16 nm is consistent with the experimentally measured spacing between the two lasing modes (14 nm).

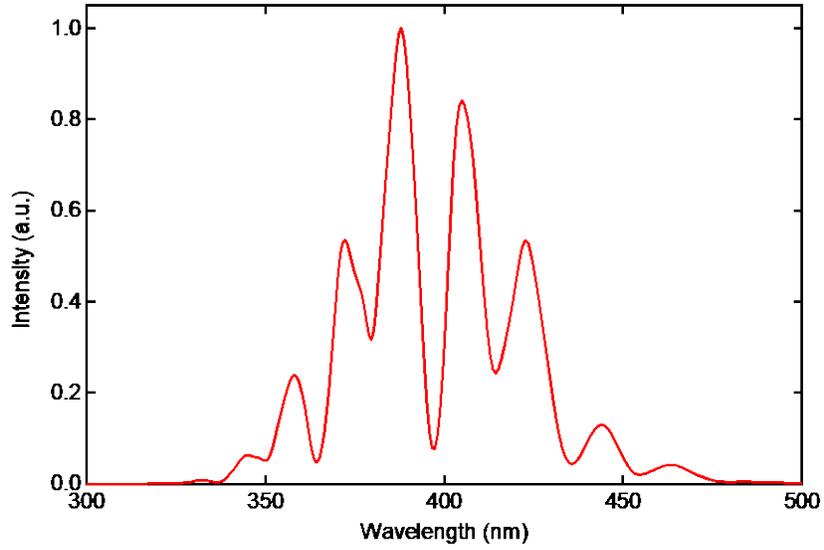

**Figure S4:** Three-dimensional finite difference time domain (FDTD) simulated spectra for a ZnO nanofin (refractive index, $n_{ZnO}$ = 2.1) of length 5 µm, width 160 nm and height 1 µm. The nanofin is placed on top of 2 µm thick GaN layer (refractive index, $n_{GaN}$ = 2.4) on a sapphire substrate ((refractive index, $n_{sapphire}$ = 1.77). The device is excited using a randomly oriented dipole placed at the interface between ZnO and GaN.

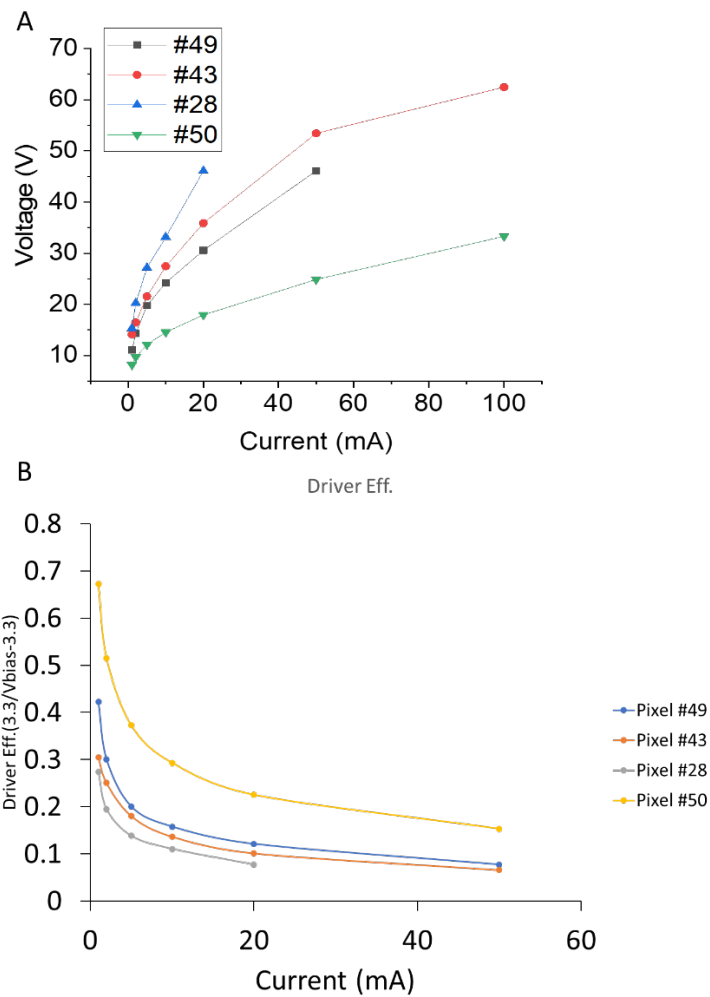

**Fig.S5**-A) Voltage-Current scans of four randomly selected pixels. B) Driving efficiencies of four different pixels that used to extract the EQE dependences.

## 5. Internal quantum efficiency for conventional and Fin LEDs using ABC model

According to the ABC model, the internal quantum efficiency is given by

$$\eta = \frac{Bn^2}{An + Bn^2 + Cn^3} = \frac{Bn/A}{1 + Bn/A + Cn^2/A} = \frac{N_p}{1 + N_p + CN_p^2 n_o^2/A} = $$
$$\frac{N_p}{1 + N_p + CN_p^2 A^2/(AB^2)} = \frac{N_p}{1 + N_p + CN_p^2 A/B^2} = \frac{N_p}{1 + N_p + c_a N_p^2}$$  (1)

Hence

$$\eta = \frac{N_p}{1 + N_p + c_a N_p^2}$$  (2)

Here $N_p = n/n_o$, $n_o = \frac{A}{B}$, and $c_a = Cn_o^2/A = CA/B^2$ is the dimensionless Auger recombination constant (see Fig. S6).

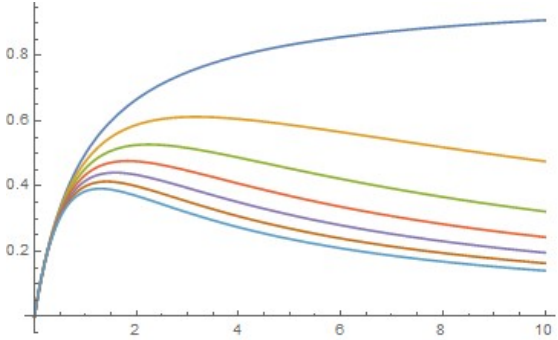

**Figure S6.** Planar LED: IQE versus $n/n_o$ for $c_a$ from 0 to 0.6

In the steady state, the current density flux $J = I/(qd)$

$$J = An + Bn^2 + Cn^3$$  (3)

Here $I$ is the current density, $q$ is the electron charge, and $d$ is the characteristic length of the recombination region.

Eq. (3) could be rewritten as

$$j = N_p + N_p^2 + c_a N_p^3$$  (4)

Here $j = J/(An_o)$. We consider two limiting cases:
(1) Low injection when the generation occurs close to the heterointerface, $d$ is independent of $J$, and we neglect the Auger recombination
(2) High injection when the Auger recombination is dominant

In the low injection case, we neglect by the Auger recombination and obtain

$$j = N_p + N_p^2$$  (5)

Solving Eq. (5) we obtain

$$N_p = \sqrt{\frac{1}{4} + j} - \frac{1}{2}$$  (6)

Hence

$$\eta_{low} = \frac{N_p}{1+N_p+c_a N_p^2} = \frac{N_p}{1+N_p} = \frac{N_p^2}{N_p+N_p^2} = \frac{N_p^2}{j} = \frac{j-\sqrt{1/4+j}+1/2}{j}, \qquad (7)$$

(see Fig. S7)

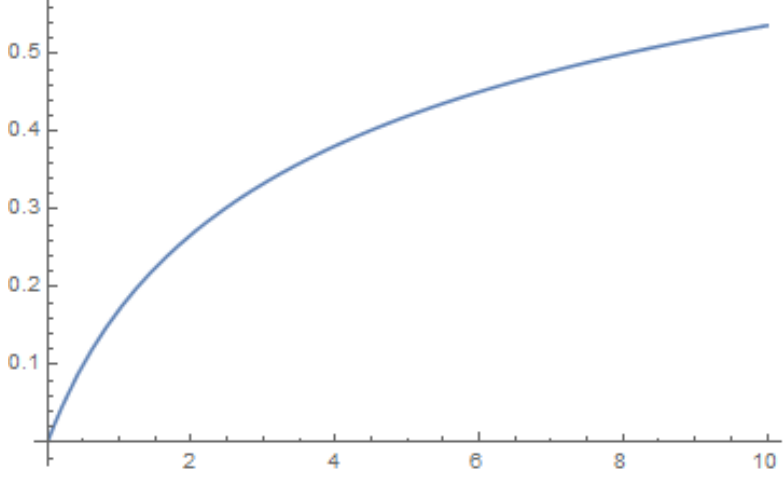

**Figure S7.** IQE versus $j$ for the low injection case.

In the high injection case, the Auger recombination is dominant, and the equation for the current flux density becomes

$$j = N_p + N_p^2 + c_a N_p^3 = c_a N_p^3 \qquad (8)$$

Hence

$$N_p = (j/c_a)^{1/3} \qquad (9)$$

$$\eta_{high} = \frac{N_p}{1+N_p+c_a N_p^2} = \frac{N_p}{c_a N_p^2} = \frac{N_p^2}{c_a N_p^3} = \frac{N_p^2}{j} = \frac{(j/c_a)^{2/3}}{j} = \frac{1}{c_a^{2/3} j^{1/3}} \qquad (10)$$

This equation applies when $c_a N_p \gg 1$ or $(jc_a^2)^{1/3} \gg 1$. The general expression for IQE could be interpolated as

$$\eta = \frac{1}{1/\eta_{high}+1/\eta_{low}} \qquad (11)$$

Assuming that $d$ is independent of $I$, we obtain the dependencies shown in Fig. S8.

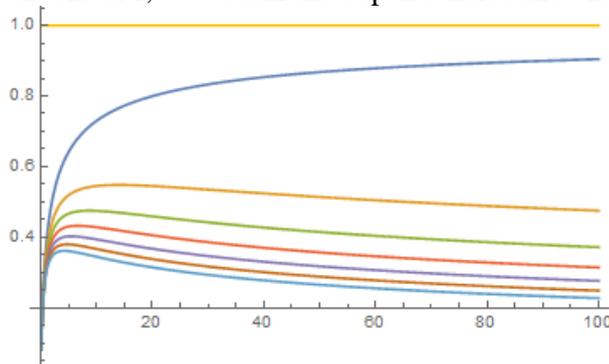

**Figure S8.** IQE vs. dimensionless current flux $j$ for conventional LEDs. The dimensionless Auger recombination constant $c_a$ varies from 0 to 0.6 (step 0.1).

Now for the case of the Fin LED we assume that at high injections the electron hole pairs spread out and the electron-hole flux gradient $j_{fin}$ saturates at the certain value $j_o$. For example, $j_{fin}$ could vary as

$$j_{fin} = j/(1+ j/j_o) \qquad (12)$$

The exact variation expression is not important for our model. What is important is that at high currents, $j_{fin} = j_o$, so that we have for the high injection case

$$\eta_{highfin} == \frac{1}{c_a^{2/3} j_o^{1/3}} \qquad (13)$$

Then using the interpolation

$$\eta = \frac{1}{1/\eta_{highfin} + 1/\eta_{low}} \qquad (14)$$

we obtain the dependencies shown in Fig. S9.

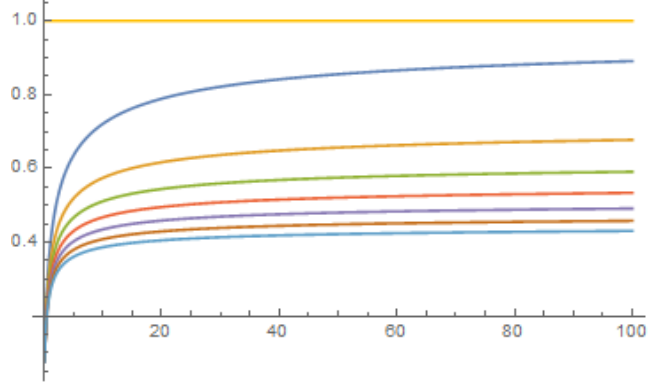

**Figure S9.** IQE vs. dimensionless current flux $j$ for Fin LED. The dimensionless Auger recombination constant $c_a$ varies from 0 to 0.6 (step 0.1). The parameter used in the calculation is the dimensionless injection current flux $j_o = 5$.